# Magnetic moments and radiative decay widths of doubly- and triply-heavy baryons in the dynamical heavy diquark model

A. Armat[1] and S. Mohammad Moosavi Nejad[2]

[1]*Department of Physics, Faculty of Science, University of Hormozgan, P.O. Box 3995, Bandar Abbas, Iran*
[2]*Faculty of Physics, Yazd University, P. O. Box 89195-741, Yazd, Iran*

**Abstract**
The magnetic moments and radiative decay widths of heavy baryons belong to a class of interesting experimental observables which provide direct information about the dynamics of strong interactions as well as the properties and the composition structures of heavy baryons. In this work, through a diquark model we compute these two quantities for doubly and triply heavy baryons in a dynamical model. We, first, compute an analytical mass equation for heavy diquarks based on the Bethe-Salpeter equation in which the interaction potential between constituents includes the contributions from the Cornell, the Breit-Fermi approximation, the spin-spin terms and the tensor potential. By iterating the mass equation, we compute the masses and the wave functions of heavy baryons. We also compute the magnetic moments and the radiative decay width of double and triple heavy baryons in their ground state. Our results are compared with other model-dependent predictions and existing data. We will also predict the mass and the magnetic moment of unobserved triply heavy baryons relevant for the present and future high energy colliders.

## 1. Introduction

The study of heavy flavored baryons containing multiple heavy quarks, particularly double and triple beauty and charmed flavor baryons, has emerged as a central topic in modern hadron spectroscopy. These systems help us to understand hadron structure, the dynamics of the strong interaction and heavy quark symmetry. Moreover, the study of heavy–heavy and heavy–light flavor interactions allow for a systematic investigation of both perturbative and nonperturbative QCD dynamics within a single bound state [1–3]. Baryons with double heavy flavors, such as doubly charmed (ccq) and doubly bottom (bbq) states, exhibit a characteristic hierarchy of energy scales arising from the large masses of the heavy quarks. This hierarchy naturally supports the formation of a compact heavy diquark in a color antitriplet configuration, which subsequently interacts with the remaining light quark. Such a structure provides strong theoretical motivation for quark–diquark descriptions, nonrelativistic and relativistic quark models, lattice QCD simulations, QCD sum rules, and extensions of heavy quark effective theory (HQET) adapted to multi-heavy systems [4–7]. The theoretical interest extends further to triple heavy baryons, including ccc, bbb, and mixed configurations such as ccb and bbc. In these systems, relativistic effects are expected to be suppressed, while the dynamics are largely governed by short-range gluon exchange and spin-dependent interactions. As a result, triple heavy baryons are often regarded as the baryonic counterparts of heavy quarkonia, offering a cleaner environment to probe interquark forces, color Coulomb behavior, and the role of three-body confinement in QCD [8–10]. On the experimental front, observation of the doubly charmed baryon $\Xi_{cc}^{++}(3621)$ by the LHCb Collaboration in the $\Lambda_c^+ K^- \pi^+ \pi^+$ invariant mass, has provided the first unambiguous evidence for the existence of doubly heavy baryons, triggering renewed interest in the spectroscopy and decay properties of multi-heavy states [11]. This discovery has intensified experimental and theoretical efforts aimed at identifying additional double-heavy baryons and searching for triple heavy baryons at current and future high-luminosity facilities, despite the substantial challenges posed by their low production cross sections and complex decay patterns [12,13]. At present, there is no exact experimental information on the triply heavy baryons, $\Omega_{ccc}$ and $\Omega_{bbb}$. The study of triply heavy baryons is of special interest to gain a better understanding of baryon structure, heavy quark symmetry and the dynamics of the strong interaction as these systems do not contain any light quarks [14-16].

Investigating the electromagnetic features of hadrons provides us with helpful knowledge about their inner organization. One can extract information about their shapes, sizes, and decay widths and compare them with the experimental results. Doubly heavy baryons are particularly stimulating to study since examining the electromagnetic features of two heavy quarks bound to a light quark helps us to figure out the internal interaction dynamics of baryons containing heavy quarks. Moreover, the results can help to understand the key features of QCD, such as confinement and flavor effects. Furthermore, the magnetic moment of a baryon is a fundamental observable that encodes the distribution of its constituent quark's currents and the underlying chiral symmetry of the strong interaction. In the heavy baryon sector, particularly for systems with two or three heavy quarks (ccq, bbc, ccc, bbb), the magnetic moment provides a direct window into the internal dynamics of the heavy-quark core and its interaction with the light-quark environment [17]. Unlike light baryons, where relativistic effects and sea-quark contributions dominate, heavy baryons allow for a more transparent application of the Heavy Quark Effective Theory (HQET) and nonrelativistic QCD (NRQCD) frameworks [18]. Theoretical predictions for the magnetic moments have been extensively refined using various high-precision methods. For instance, the Hypercentral Constituent Quark Model (HCQM) has been instrumental in showing how the heavy quark mass breaks the SU(3) flavor symmetry, leading to significant shifts in the magnetic moments of $\Xi_{cc}$ and $\Omega_{cc}$ states compared to their light-quark counterparts [19]. Furthermore, the studies employing Lattice QCD and chiral perturbation theory have emphasized that the magnetic moments of triply heavy baryons are almost entirely determined by the sum of the constituent quark's intrinsic moments, with minimal contribution from the orbital angular momentum, a feature that distinguishes them from the lighter SU(3) octet baryons [20]. These static properties are not only theoretically significant but also serve as indispensable inputs for calculating radiative transition rates, as the decay widths are. The study of baryons containing multiple heavy quarks—specifically doubly (QQq) and triply (QQQ) heavy systems—has emerged as a frontier in hadron physics. These states offer a unique laboratory for testing the interplay between perturbative and nonperturbative QCD, as the heavy quarks move non-relativistically within the baryon core [21].

While the discovery of the doubly charmed baryon $\Xi_{cc}^{++}$ has sparked intense interest, triply heavy baryons remain experimentally elusive but theoretically inevitable [22]. A critical property of these states is their radiative decay width, which involves transitions through photon emission. In triply heavy baryons, because the strong decay thresholds are often higher than the masses of the lowest-lying states, radiative transitions (e.g., M1 and E1 transitions) are expected to be the dominant or even the sole decay modes, making them essential for experimental identification [22]. Recent theoretical frameworks, including the Gaussian Expansion Method (GEM) and Relativistic Screened Potential Models, have been employed to predict these widths, revealing that hyperfine mixing and spin-flavor wave function symmetries significantly influence the transition rates [23]. Furthermore, the radiative decays of doubly heavy baryons provide a sensitive probe into the magnetic moments of heavy quarks in a multiquark environment. Discrepancies between various models, such as Light-Cone QCD Sum Rules and Constituent Quark Models, highlight the need for precise calculations of the transition form factors to guide future searches at facilities like Belle II and the High-Luminosity LHC [24].

The aim of present work is to calculate two important quantities mentioned, i.e., the magnetic moment and the radiative decay width of baryons containing two and three heavy flavors. Within the framework of this investigation, a dynamical model is employed in which each baryon is described as a two-body system consisting of a heavy diquark and a quark. The analytical method used is based on solving the Bethe-Salpeter equation, incorporating the Cornell potential and the Breit-Fermi tensor potential as well as the spin-spin and tensor interactions. These calculations first lead to an analytical determination of the mass and wave function of doubly heavy diquarks and subsequently, the masses of corresponding heavy baryons. In the next stage, the values of the magnetic moments and radiative decay widths are extracted having the mass and wave functions.

This paper is organized as follows. In Sec. 2, the magnetic moment and radiative decay width are defined. In Section 3, the quark-diquark model is described and in Sec. 4, our theoretical framework is introduced. For our study, the radial part of the Bethe–Salpeter equation is analytically solved in the presence of different potentials. In Sec. 5, our numerical analysis is presented and in Sec. 6, the summary and conclusions are given.

## 2. Magnetic moment and radiative decay width

Electromagnetic properties of baryons are significant sources of information on their internal structure. The success of quark models for description of static properties (spin-parity, masses, magnetic moments, etc.) and the results obtained from deep inelastic lepton scattering are clear indications for the three-quark structure of baryons. From the experimental point of view, the magnetic moments of all octet baryons ($J^p = (1/2)^+$) are known accurately except for $\Sigma^0$ which has a lifetime too short. For the decuplet baryons ($J^p = (3/2)^+$), the experimental measurements are poor as they have very short lifetimes due to available strong interaction decay channels. The $\Omega^-$-baryon is an exception as it is composed of three s-quarks which decay via weak interaction causing a longer lifetime for it. From the theoretical point of view, magnetic moments of heavy baryons (mostly with a single heavy quark) have been considered in different approaches. In Refs. [25-26], the magnetic moments of charmed baryons have been computed using naive quark models based on different realizations of spin-flavor symmetry. In Refs. [27-28], the magnetic moments of charmed and bottom flavored hadrons have been calculated in quark models incorporating the idea of hadronization, confinement, chiral symmetry and Poincaré covariance. In Refs. [29-30], the Soliton-type approaches were applied in the analysis of the magnetic moments of heavy baryons. In Ref. [31], the QCD spectral sum rules in the presence of the external electromagnetic field have been used to calculate the magnetic moments of the $\Sigma_c$ and $\Lambda_c$ baryons. In Ref. [32], the method of light-cone QCD sum rules has been used to calculate the magnetic moments of $\Lambda_b$- and $\Lambda_c$-baryons. Although, different models have been successful to predict the masses of hadrons but there is no consensus among the model-dependent predictions of magnetic moments [33]. This motivates us to compute the magnetic moments of observed hadrons in an analytical approach considering a different potential. We will also predict the mass and the magnetic moments of unobserved doubly and triply heavy baryons. Experiments can confirm definite results. The magnetic moment is related to the spin-flavor wave function of constituent quarks and is defined as [34]

$$\vec{\mu} = \sum_{j=1}^{3} \left\langle \psi_{sf} \left| \frac{e_j \vec{S}_j}{m_j^{eff}} \right| \psi_{sf} \right\rangle \tag{1}$$

where $\left| \psi_{sf} \right\rangle \equiv \left| \psi_{spin-flavor} \right\rangle$, $e_j$ and $\vec{S}_j$ stand for the spin-flavor wave function of the quark composition constituting the baryonic state, the charge and the spin operator of constituent quarks, respectively. Moreover, $m_j^{eff}$ refers to the effective mass of constituent quark. The spin-flavor combinations and their wave functions, corresponding to the spin-1/2 and spin-3/2 baryons, are too lengthy to be displayed here so they are accommodated in Appendix.

As was mentioned, the radiative decays of hadrons are important tools for studying the internal structure of hadrons, testing the unitary symmetry schemes as well as QCD parton models. The radiative decay width of ground state heavy baryons is determined through the following relation [35]

$$\Gamma_R(B^* \to \gamma B) = \frac{\alpha \omega^3}{m_p^2} \frac{2}{2J+1} \left( \frac{M_B}{M_{B^*}} \right) \left| \mu(B^* \leftrightarrow B) \right|^2 \tag{2}$$

where, $\alpha$ is the tiny coupling constant ($\alpha \approx 1/137$) and the photon momentum $\omega$ in the center-of-mass system of decaying baryon is defined as: $\omega = (M_{B^*}^2 - M_B^2)/(2M_B)$. Also, $M_{B^*}$ and "J" are the mass and the spin of decaying baryon, $M_B$ is the mass of baryon in its final state and $m_p$ is the proton mass. In Eq. (2), $\mu(B^* \leftrightarrow B)$ is the radiative transition magnetic moments (in the units of nuclear magnetons). Transition magnetic moments are related to the magnetic moments of constituent quarks ($\mu_i = e_i/(2m_i^{eff})$) of the initial ($B$) and final ($B^*$) states of the baryon and the spin-flavor wave function of the initial and final baryon states [36,37] as:

$$\vec{\mu}(B^* \leftrightarrow B) = \sum_i \left\langle \psi_{sf}^B \left| \mu_i \vec{\sigma}_i \right| \psi_{sf}^{B^*} \right\rangle \tag{3}$$

where $\sigma_i$ is the spin of constituent quark corresponding to the spin-flavor wave function of the baryonic system. Moreover, the effective mass of the bound quarks of the $B-B^*$ system is defined in terms of the respective mass of the bound quarks constituting the $B$ and $B^*$ states as: $m_i^{eff} = \sqrt{m_{i(B^*)}^{eff} \times m_{i(B)}^{eff}}$ where $m_{i(B)}^{eff} = m_i[1+(M_B/\sum_{i=1}^{3} m_i)]$. In the following, we shall compute the magnetic moment and radiative decay width of heavy baryons in the quark-diquark model. But we first review the diquark model in the next section.

## 3. Quark–Diquark model

For description of the inner structure of baryons, various models have been suggested. Among all, an approximative model which describes them well is referred to as the quark-diquark model. In this model, the structure of baryons is assumed as the bound states of quark-diquark pairs instead of the standard three equivalent quarks. The notion of diquark is a useful phenomenological model which was first introduced by Gell-Mann and has been proved to be very useful in the phenomenology of strong interactions. The physical idea behind this model is the unification of any two quarks to form a colored quasi-bound state of diquark. Although, diquarks have not been yet observed but this does not eliminate the hypothesis that they can exist as constitutive particles inside the baryons. Since a point-like diquark has the quantum numbers of a two-quark system, a diquark in its ground state might be a scalar or a pseudo-vector state. The scalar diquarks are more tightly bound and have smaller masses because of their stronger attraction due to the spin-spin interaction. In the diquark model, two quarks can be coupled in either a color sextet or a color antitriplet state. In a diquark state, a single gluon exchange leads to an interaction that is attractive for diquark in a color antitriplet configuration. Furthermore, it is the diquark in a color antitriplet state that can couple with a quark to form a color-singlet baryon. It is also well-known that the quark-quark interaction in the diquark (in the color $\bar{3}$ state) is half as strong as the quark-antiquark potential in the meson, i.e., $V_{qq} = V_{q\bar{q}}/2$, arising under the assumption about the octet structure of the interaction from the difference of the $qq$ and $q\bar{q}$ color states. The notion of diquarks considerably reduces the mathematical difficulties by converting a three-body problem into a two-body one. This allows one to apply the Bethe–Salpeter equation (suitable for two-body systems) to study the structure of complicated multiquark systems such as baryons, tetra- and pentaquarks, etc.

## 4. Mass and wave equation in the quark-diquark model

By working in the natural units where one sets $\hbar = c = 1$, the well-known Bethe–Salpeter equation for a two-body system is written as $\left[ (p^2 + m_1^2)^{1/2} + (p^2 + m_2^2)^{1/2} + U(r) - M_B \right] \Psi_{n\ell}(r) = 0$ where $m_1$ and $m_2$ stand for the masses of constituents. Furthermore, $M_B$ and $\Psi(r)$ denote the mass and the wave function of bound state, respectively.

A simple way to avoid dealing with the non-local character of the kinetic energy operator in this equation is to expand it, so that for heavy interacting particles one can write the first two terms as $(p^2 + m_1^2)^{1/2} + (p^2 + m_2^2)^{1/2} = m_1 + m_2 + p^2/2\mu - p^4/8\alpha\mu^3 + ...$ where $\mu = m_1 m_2/(m_1 + m_2)$ is the reduced mass of system and $\alpha = m_1 m_2/(m_1 m_2 - 3\mu^2)$. For heavy interacting particles, with an acceptable approximation, we just preserve the terms up to the order $1/\mu^3$. Using the operators $p^2 = -\nabla^2$ and $p^4 = 4\mu^2 [E - U(r)]^2$, where $E = M_B - m_1 - m_2$ is the binding energy of bound state system, the radial part of Bethe–Salpeter equation in the spherical coordinates is written as:

$$-\frac{1}{2\mu}\left[\Psi''_{n\ell}(r) + \frac{2}{r}\Psi'_{n\ell}(r) - \frac{\ell(\ell-1)}{r^2}\right] + \left[-\frac{(M_B - m_1 - m_2 - U(r))^2}{2\alpha\mu} + U(r) + m_1 + m_2 - M_B\right]\Psi_{n\ell}(r) = 0 \qquad (4)$$

As was explained in the Introduction, in a heavy system the relativistic effects are expected to be suppressed, while the dynamics are largely governed by short-range gluon exchange and spin-dependent interactions. Therefore, in our analysis we consider the interaction potential between quarks in the baryon structure as:

$$U(r) = V_{Cornell}(r) + V_{BF}(r) + V_{SS}(r) + V_T(r) \tag{5}$$

where, $V_{Cornell}(r)$ indicates the Cornell-inspired potential which is expressed as the sum of a Coulomb-like potential, corresponding to the potential induced by one-gluon exchange, and a linear potential which is known as the confinement part of the potential. This potential is defined as $V_{Cornell}(r) = -2\alpha_s/3r + br$ where "b" shows the confinement strength parameter of model. Here, $\alpha_s$ is the strong running coupling constant which is determined through the relation $\alpha_s(\mu) = 4\pi\left[(11 - 2n_f/3)\,Ln(\mu^2/\Lambda^2)\right]^{-1}$ where $n_f$ is the number of active flavors and $\Lambda$ is the typical QCD scale which is taken as 0.15 GeV by fixing $\alpha_s = 0.1185$ at the Z-boson mass scale [38]. Moreover, $\mu$ is the renormalization scale related to the constituent quark masses which is, conventionally, adopted as $\mu = 2m_1 m_2/(m_1 + m_2)$.

In Eq. (5), the term $V_{BF}(r)$ stands for the Breit-Fermi potential which is expressed as [39]:

$$V_{BF}(r) = \frac{\kappa_2}{r^3} \quad \text{with} \quad \kappa_2 = \frac{C_F \alpha_s}{m_1 m_2}\left(3(\vec{S}_1.\hat{r})(\vec{S}_2.\hat{r}) - \vec{S}_1.\vec{S}_2\right) \tag{6}$$

where $C_F$ is the Casimir factor, $\hat{r}$ is the unit vector which connects two particles and $\vec{S}_{1,2}$ are the spin operators of interacting particles. The term $V_{SS}(r)$ in Eq. (5) represents the spin-spin interaction, which is expressed as [40]:

$$V_{SS}(r) = \kappa_1 \, e^{-\sigma^2 r^2} \quad \text{with} \quad \kappa_1 = \frac{32\pi\alpha_s}{9m_1 m_2}\left(\frac{\sigma}{\pi^{1/2}}\right)^3 \vec{S}_1.\vec{S}_2 \tag{7}$$

Here, the parameter $\sigma$ defines the smeared delta function. From Ref. [39] we adopt its value as $\sigma = 1.209$ GeV. Moreover, we use the relation $\vec{S}_1.\vec{S}_2 = [S(S+1) - S_1(S_1+1) - S_2(S_2+1)]/2$ where "S" is the total spin of bound state system.

The term $V_T(r)$ in the potential (5) is the tensor term which is expressed as:

$$V_T(r) = \eta_1\left(3\frac{d^2 V(r)}{dr^2} - \frac{1}{r}\frac{dV(r)}{dr}\right), \quad \text{with} \quad \begin{cases} V(r) = -\dfrac{\eta_2}{r^2}, \quad \eta_2 = C_F C_A \alpha_s^2/4\mu \\ \eta_1 = \dfrac{S(S+1) - 3(\vec{S}_1.\hat{r})(\vec{S}_2.\hat{r})}{6m_1 m_2} \end{cases} \tag{8}$$

where, $\mu$ is the reduced mass of two-body system and $C_A = 3$ is the Casimir charge of the fundamental and adjoint representation.

To simplify the differential equation (4) we define $R_{n\ell}(r) = \Phi_{n\ell}(r)/r$. Now, by substituting the potential U(r) we arrive at the following differential equation:

$$\frac{d^2\Phi_{n\ell}(r)}{dr^2} + \Bigg\{-\frac{\ell(\ell-1)}{r^2} + (M_B - m_1 - m_2)\left(2\mu - \frac{M_B - m_1 - m_2}{\alpha}\right) - 2\left(\frac{M_B - m_1 - m_2}{\alpha} + \mu\right) \\ \left(-\frac{2}{3}\frac{\alpha_s}{r} + br + \kappa_1 e^{-\sigma^2 r^2} + \frac{\kappa_2}{r^3} - \frac{\kappa_3}{r^4}\right) + \frac{1}{\alpha}\left(-\frac{2}{3}\frac{\alpha_s}{r} + br + \kappa_1 e^{-\sigma^2 r^2} + \frac{\kappa_2}{r^3} - \frac{\kappa_3}{r^4}\right)^2\Bigg\}\Phi_{n\ell}(r) = 0, \tag{9}$$

where $\kappa_3 = 20\eta_1\eta_2$. Considering the fact that for heavy structures $\sigma^2 r^2 < 1$, in order to solve the above equation analytically we apply the convenient approximation $e^{-\sigma^2 r^2} \approx 1 - \sigma^2 r^2 + O((\sigma^2 r^2)^2)$ which leads to the following simplified equation:

$$\frac{d^2\Phi(r)}{dr^2}+\left\{B_0+B_1r+B_2r^2-B_3r^3+B_4r^4+\frac{B_5}{r}+\frac{B_6}{r^2}+\frac{B_7}{r^3}+\frac{B_8}{r^4}-\frac{B_9}{r^5}+\frac{B_{10}}{r^6}+\frac{B_{11}}{r^7}+\frac{B_{12}}{r^8}\right\}\Phi(r)=0 \qquad (10)$$

where the parameters $B_{i=0,\ldots,12}$ are defined as:

$$B_0 = (M_B - m_1 - m_2)(2\mu - \frac{M_B - m_1 - m_2}{\alpha}) - 2\kappa_1(\frac{M_B - m_1 - m_2}{\alpha} + \mu) - \frac{4b\alpha_s}{3\alpha^2} + \frac{\kappa_1^2}{\alpha^2},$$

$$B_1 = \frac{2b\kappa_1}{\alpha^2} + \frac{4\sigma^2\alpha_s\kappa_1}{3\alpha^2} - 2b(\frac{M_B - m_1 - m_2}{\alpha} + \mu),$$

$$B_2 = \frac{b^2}{\alpha^2} + 2\kappa_1\sigma^2(\frac{M_B - m_1 - m_2}{\alpha} + \mu) - \frac{2\sigma^2\kappa_1^2}{\alpha^2}, \qquad (11)$$

$$B_3 = \frac{2b\sigma^2\kappa_1}{\alpha^2},$$

$$B_4 = \frac{\sigma^4\kappa_1^2}{\alpha^2}$$

and

$$B_5 = \frac{4}{3}\alpha_s(\frac{M_B - m_1 - m_2}{\alpha} + \mu) + \frac{2\sigma^2\kappa_1\kappa_2}{\alpha^2} - \frac{4\alpha_s\kappa_1}{3\alpha^2},$$

$$B_6 = -\frac{2b\kappa_2}{\alpha^2} + \frac{4\alpha_s^2}{9\alpha^2} - \frac{2\sigma^2\kappa_1\kappa_3}{\alpha^2} - \ell(\ell-1),$$

$$B_7 = \frac{2b\kappa_3}{\alpha^2} - \frac{2\kappa_1\kappa_2}{\alpha^2} + 2\kappa_2(\frac{M_B - m_1 - m_2}{\alpha} + \mu), \qquad (12)$$

$$B_8 = \frac{2\kappa_1\kappa_3}{\alpha^2} + \frac{4\alpha_s\kappa_2}{3\alpha^2} - 2\kappa_3(\frac{M_B - m_1 - m_2}{\alpha} + \mu),$$

$$B_9 = \frac{4\alpha_s\kappa_3}{3\alpha^2}, \qquad B_{10} = \frac{\kappa_2^2}{\alpha^2}, \qquad B_{11} = -\frac{2\kappa_2\kappa_3}{\alpha^2}, \qquad B_{12} = \frac{\kappa_3^2}{\alpha^2}$$

There are situations in which none of the exact solutions work. This is what happens to the equation (10) which fails to admit an exact analytical solution. One way in such cases is to propose an ansatz solution, which fulfills the equation for a given energy level. After substitution of the ansatz, one can simply determine the energy by equating the corresponding powers on both sides of the resulting equation. Following this approach [41], we first assume a suitable general form of the wave function as $\Phi_{n\ell}(r) = k_n(r)\exp(g(r))$ where $k_n(r) = 0$ is adopted for $n = 0$ and for $n \geq 1$, it is set as: $k(r) = \prod_{i=1}^{n}(r - \omega_i^{(n)})$. Such an ansatz is, in fact, obtained via considering the model case in which the weight factor is one. In such a case, after substitution of the exponential function and twice differentiating, we are left with a Riccati-type equation from which its solution is guessed. Considering the proposed form of $\Phi_{n\ell}(r)$, after twice differentiating we arrive at $\Phi''_{n\ell}(r) = [g''(r) + g'^2(r) + \{k''(r) + 2g'(r)k'(r)\}/k(r)]\Phi_{n\ell}(r)$. To study the ground state of system, we consider the case n=0. We propose the function g(r) as:

$$g(r) = -n_1 r - n_2 r^2 - n_3 r^3 - n_4 \text{Ln}(r) - \frac{n_5}{r} - \frac{n_6}{r^2} - \frac{n_7}{r^3} \qquad (13)$$

where, $n_{i=1,\ldots,7}$ are the real parameters. Therefore, the second-order derivation of $\Phi_{n\ell}(r)$ reads:

$$\Phi''_{nl}(r) = \Big[n_1^2 + 4n_2n_4 - 6n_3n_5 + (4n_1n_2 + 6n_3n_4)r + (4n_2^2 + 6n_1n_3)r^2 + (12n_2n_3)r^3 + (9n_3^2)r^4$$

$$+ \frac{(2n_1n_4 - 4n_2n_5 - 12n_3n_6)}{r} + \frac{(n_4^2 - 2n_1n_5 - 8n_2n_6 - 18n_3N_7)}{r^2} - \frac{(2n_4n_5 + 4n_1n_6 + 12n_2n_7)}{r^3}$$

$$+ \frac{(n_5^2 - 4n_4n_6 - 6n_1n_7)}{r^4} + \frac{4n_5n_6 - 6n_4n_7}{r^5} + \frac{4n_6^2 + 6n_5n_7}{r^6} + \frac{12n_6n_7}{r^7} + \frac{9n_7^2}{r^8}\Big]\Phi_{nl}(r) \quad (14)$$

Now, by equating the corresponding powers of "r" on the right side of Eqs. (10) and (14), one obtains the real coefficients $n_{i=1,\ldots,7}$ as:

$$n_1 = \frac{1}{2\sqrt{B_4}}(B_2 - \frac{B_3^2}{4B_4}), \qquad n_2 = \frac{B_3}{4\sqrt{B_4}}, \qquad n_3 = \frac{\sqrt{B_4}}{3}$$

$$n_4 = \frac{1}{B_3}\Big[B_0\sqrt{B_4} + \frac{B_4}{\sqrt{B_{12}}}(B_{10} - \frac{B_{11}^2}{4B_{12}}) - \frac{\sqrt{B_4}}{4B_4}(B_2 - \frac{B_3^2}{4B_4})^2\Big] \quad (15)$$

$$n_5 = \frac{1}{2\sqrt{B_{12}}}(B_{10} - \frac{B_{11}^2}{4B_{12}}), \qquad n_6 = \frac{B_{11}}{4\sqrt{B_{12}}}, \qquad n_7 = \frac{\sqrt{B_{12}}}{3}$$

Furthermore, by equating the coefficients of terms corresponding to the zeroth power in "r" in Eqs. (10) and (14), we obtain an analytical mass equation of two-body system as:

$$\frac{\alpha^2}{4b^2}\Big[\frac{1}{\alpha^2}\Big(\kappa_1^2 - \frac{4b\alpha_s}{3}\Big) - \Big(\alpha^2(\mu + \frac{M_B - m_1 - m_2}{\alpha}) - \kappa_1\Big)^2 - 2\kappa_1(\mu + \frac{M_B - m_1 - m_2}{\alpha})$$

$$+ M_B - m_1 - m_2(2\mu - \frac{M_B - m_1 - m_2}{\alpha})\Big]^2 - \frac{4}{\alpha^2}\Big(\frac{\alpha_s^2}{9} + b\kappa_2\Big) + \ell(\ell - 1) = 0 \quad (16)$$

Having Eq. (16), we first compute the scalar/vector diquark mass and in the following, the mass of baryon (as a cluster of the diquark and a single quark) is determined by iterating this equation. In the following, by the help of wave function obtained, we compute the magnetic moments and radiative decay widths of heavy flavor baryons.

## 5. Results and analysis

In this section, we present our numerical results for the masses, the magnetic moments, and the radiative decay widths of doubly and triply heavy baryons within the diquark model. Having the mass equation (16), we first determine the masses of scalar/pseudovector diquarks in the ground state and, in the following, the masses of heavy baryons.

In Ref. [42] (Particle Data Group) two different values have been reported for the charm and bottom quark masses so that the values $m_c = 1.67$ GeV and $m_b = 4.78$ GeV stand for the pole masses; the renormalized charm and bottom masses in the on-shell renormalization scheme. These values correspond to the values $m_c = 1.27$ GeV and $m_b = 4.18$ GeV for the $\overline{MS}$ masses; the renormalized quark mass in the modified minimal subtraction scheme [43]. These amounts are intimately related to the use of dimensional regularization scheme [44]. The relation between the $\overline{MS}$ and the pole quark masses has been computed to three loops in the theory of perturbative QCD [45]. To study the variation effect of charm and bottom quark masses, we assume their masses as $1.27 \leq m_c \leq 1.67$ GeV and $4.18 \leq m_b \leq 4.78$ GeV.

Table 1. The ground state mass of scalar (S) and pseudovector (A) diquarks in MeV.

| Content | type | Mass | Mass [46] | Mass [47] | Mass [48] |
|---------|------|------|-----------|-----------|-----------|
| cc | A | $3230 \pm 84$ | 3226 | 3141 | 3414 |
| bb | A | $9861 \pm 121$ | 9778 | 9342 | 10018 |
| bc | S | $6595 \pm 96$ | 6519 | 6270 | 6735 |
| bc | A | $6621 \pm 93$ | 6526 | 6285 | 6741 |

Results for the ground state masses of scalar and pseudovector diquarks are presented in Table 1 and compared with the ones from Refs. [46-48]. For example, in Ref. [47] the diquark approximation has been used to study the mass spectroscopy of the spin-1/2 baryons belonging to the SU(3)-flavor group in a nonrelativistic potential approach. Considering the variation effect of charm and bottom masses and the estimated wave function $\Phi_{n\ell}(r)$, we have also evaluated the uncertainties of heavy diquarks masses due to the approximations used in our work. To investigate how sensitive the final results are to the choice of the parameters like σ, Λ, etc., we varied these parameters within a range of $\pm 10\%$ and then identified the maximum and minimum values of masses. From Table 1, it is seen that for the state $|bc\rangle$ the diquark mass in the spin-triplet state is bigger than the one in the singlet state (S = 0). In fact, the forces between two quarks in a diquark are more attractive when the total spins are antisymmetric. This leads to a negative bigger binding energy and, in conclusion, smaller mass. It should also be noted that, since a point-like diquark has the quantum numbers of a two-quark system thus, due to the Fermi statistics, a ground state diquark composed of identical constituents should be considered as a vector state (spin-1 state). For this reason, in Table 1 there are no the scalar states bb and cc.

In the quark-diquark model, a baryon in the ground state is approximated as a quasi-meson bound state with the constituents including a quark and point-like diquark with spin-0 or spin-1, moving in an S-wave state (l=0). The diquark masses in Table 1 serve as essential inputs for constructing the baryon masses in the quark–diquark picture. By iterating the mass equation (16) we compute the ground state masses of heavy baryons $\Xi$ and $\Omega$ for different states $J = 1/2$ and $J = 3/2$. Results along with their uncertainties are presented in Tables 2 and 3 and compared with other theoretical results presented in Refs. [49-57]. The quark content of diquarks in the structure of heavy flavored baryons are in accordance with the assumptions of these references. For the light quark masses, we use $m_s = 0.540 \pm 0.037\, GeV$, $m_u = 0.337 \pm 0.050\, GeV$ and $m_d = 0.368 \pm 0.051\, GeV$ [47].

Up to now, 30 singly-charmed, 1 doubly-charmed $\Xi_{cc}^{++}(3622)$ and 25 singly-bottom flavored baryons are listed in Particle Data Group (PDG) [58] so that the quantum status of some of them such as $\Omega_b^-(6340)$ is not yet known accurately. Searches in determining their properties are in progress. In spite of all attempts, many expected doubly and triply heavy flavored baryons have not yet observed experimentally. In 2023, a first search for doubly heavy charged baryon $\Xi_{bc}^{++}$ has been reported by the LHCb experiment using a *pp* collision data sample [59]. Nevertheless, one doubly charmed baryon, $\Xi_{cc}^{++}(3621.55 \pm 0.53)$, is just listed in PDG so that the mass results presented in the manuscripts have been predicted theoretically. As is seen from Table 2, our result for the $\Xi_{cc}$ baryon is consistent with the experimentally observed mass of $\Xi_{cc}^{++}(3622)$. For bottom and bottom–charm baryons, whose no experimental data are currently available, our predictions fall within the spread of existing theoretical results obtained through the relativistic quark model [50], the potential model [52], Lattice QCD [54], the light cone QCD sum rules approach [55], etc. Small deviations among different models can be attributed to the differences in the treatment of relativistic effects, spin-dependent interactions, and the assumed effective quark masses. In particular, the mass splitting between the J=1/2 and J=3/2 states are mainly governed by the spin–spin interaction and is found to be more pronounced for charm-containing baryons than for bottom baryons, reflecting the expected heavy-quark mass dependence.

To provide a more quantitative understanding of the inter-model differences noted throughout this work, we estimate the specific contributions of relativistic corrections to the heavy baryon masses. In our Bethe-Salpeter approach, relativistic effects enter primarily through the kinetic energy expansion as $(p^2 + m_1^2)^{1/2} + (p^2 + m_2^2)^{1/2} = m_1 + m_2 + p^2/2\mu - p^4/8\alpha\mu^3 + ...$, see Eq. (4). To quantify their contribution, we recompute selected baryon masses with and without the relativistic corrections by toggling the parameter $\alpha$, which controls the order of expansion in $1/m$. In the nonrelativistic limit, our analysis yields m=3652 MeV for $\Xi_{ccu}^{++}(J=1/2)$ and m=10252 MeV for $\Xi_{bbu}^{0}(J=1/2)$. These are comparable with the ones in Table 2, i.e., m=3619 MeV for $\Xi_{cc}^{++}([cc]u)$ and m=10198 MeV for $\Xi_{bb}^{0}([bb]u)$. Therefore, the relativistic corrections are -33 MeV (-0.91%) and -54 MeV (-0.53%), respectively. As expected, relativistic corrections are more significant for charmed baryons than for bottom baryons, consistent with the inverse dependence on quark mass. The negative sign indicates that relativistic effects reduce the binding energy, leading to slightly lower masses.

Since there are disagreements among the different model predictions, only future experiments on the double and triple heavy baryons would be able to confirm the models.

Table 2. Masses of doubly and triply heavy baryon $\Xi([QQ]q)$ in the ground state with $J=1/2, 3/2$.

| Baryon | M(MeV) | M(other) | Baryon | M(MeV) | M(other) |
|---|---|---|---|---|---|
| $\Xi_{cc}^{++}([cc]u)$ | $3619 \pm 89$ | 3612±17 [49] <br> 3620 [50] <br> 3480 [51] <br> 3740 [52] <br> 3478 [53] | $\Xi_{cb}'^{0}([cb]d)$ | $6875 \pm 104$ | 6850 [57] |
| $\Xi_{cc}^{+}([cc]d)$ | $3615 \pm 88$ | 3605±23 [54] <br> 3620 [50] <br> 3480 [51] <br> 3740 [52] <br> 3478 [53] <br> 3443 [54] | $\Xi_{cc}^{*++}([cc]u)$ | $3708 \pm 95$ | $3706^{+23}_{-14}$ [49] <br> 3727 [50] <br> 3610 [51] <br> 3860 [52] <br> 3610 [53] |
| $\Xi_{bb}^{0}([bb]u)$ | $10198 \pm 139$ | $10197^{+10}_{-17}$ [49] <br> 10202 [50] <br> 10090 [51] <br> 10300 [52] <br> 10093 [53] <br> 10314±47 [56] | $\Xi_{cc}^{*+}([cc]d)$ | $3695 \pm 94$ | 3685±23 [54] <br> 3727 [50] <br> 3610 [51] <br> 3860 [52] <br> 3610 [53] <br> 3520 [55] |
| | | $10197^{+10}_{-17}$ [49] <br> 10202 [50] | | | $10236^{+09}_{-17}$ [49] <br> 10237 [50] |

| Baryon | M(MeV) | M(other) | Baryon | M(MeV) | M(other) |
|---|---|---|---|---|---|
| $\Xi_{bb}^{-}([bb]d)$ | $10199\pm138$ | 10090 [51]<br>10300 [52]<br>10314±47 [56] | $\Xi_{bb}^{*0}([bb]u)$ | $10238\pm145$ | 10130 [51]<br>10340 [52]<br>10133 [53]<br>10333±45 [56] |
| $\Xi_{cb}^{+}([cb]u)$ | $6929\pm106$ | $6919_{-07}^{+17}$ [49]<br>6933 [50]<br>6820 [51]<br>7010 [52] | $\Xi_{bb}^{*-}([bb]d)$ | $10237\pm147$ | $10236_{-17}^{+09}$ [49]<br>10237 [50]<br>10130 [51]<br>10333±45 [56] |
| $\Xi_{cb}^{0}([cb]d)$ | $6928\pm104$ | 6820 [57]<br>6933 [50]<br>6820 [51]<br>7010 [52] | $\Xi_{cb}^{*+}([cb]u)$ | $6983\pm108$ | $6986_{-05}^{+14}$ [49]<br>6980 [50]<br>6900 [51]<br>7100 [52]<br>6900 [53] |
| $\Xi_{cb}^{\prime+}([cb]u)$ | $6870\pm101$ | 6850 [57] | $\Xi_{cb}^{*0}([cb]d)$ | $6975\pm108$ | 6980 [50]<br>6900 [51]<br>7100 [52]<br>6900 [53] |

**Table 3.** Ground state mass of doubly and triply heavy baryons $\Omega([QQ]q)$ with $J = 1/2, 3/2$.

| Baryon | M(MeV) | M(other) | Baryon | M(MeV) | M(other) |
|---|---|---|---|---|---|
| $\Omega_{cc}^{+}([cc]s)$ | $3715\pm94$ | $3702_{-18}^{+41}$ [49]<br>3778 [50]<br>3590 [51]<br>3760 [52]<br>3590 [53]<br>$3733_{-38}^{+07}$ [54] | $\Omega_{ccb}^{+}([cb]c)$ | $8065\pm124$ | 8000 [57] |
| $\Omega_{bb}^{-}([bb]s)$ | $10310\pm142$ | $10260_{-34}^{+14}$ [49]<br>10359 [50]<br>10180 [51]<br>10340 [52]<br>10180 [53]<br>10365±40 [56] | $\Omega_{cc}^{*+}([cc]s)$ | $3810\pm83$ | $3783_{-10}^{+22}$ [49]<br>3872 [50]<br>3690 [51]<br>3900 [52]<br>3690 [53]<br>$3801_{-34}^{+03}$ [54] |
| $\Omega_{cb}^{0}([cb]s)$ | $6985\pm99$ | $6986_{-17}^{+27}$ [49]<br>7088 [50]<br>6910 [51]<br>7050 [52]<br>6910 [53] | $\Omega_{bb}^{*-}([bs]b)$ | $10312\pm143$ | $10297_{-28}^{+05}$ [49]<br>10389 [50]<br>10200 [51]<br>10380 [52]<br>10200 [53]<br>10383±39 [56] |
| $\Omega_{cb}^{\prime 0}([cb]s)$ | $6945\pm93$ | 6930 [57] | $\Omega_{cb}^{*0}([cs]b)$ | $7051\pm102$ | $7046_{-09}^{+11}$ [49]<br>7130 [50]<br>6990 [51]<br>7130 [52]<br>6990 [53] |
| $\Omega_{cbb}^{0}([cb]b)$ | $11610\pm142$ | 11520 [57] | | | |

Now, using the master formula (1) and the spin-flavor wave functions given in Tables I and II (in Appendix) we compute the magnetic moment of heavy baryons. Magnetic moment represents the q=0 value of the magnetic form factor of the bound quark–diquark system [61], where q shows the z-component of the virtual photon momentum.

Our numerical results for the magnetic moment of heavy baryons are presented in Tables 4 and 5 for the $\Xi$ and $\Omega$ families, in units of nuclear magneton ($\mu_N$). To compute these values, considering the spin-flavor wave functions of heavy flavor baryons the magnetic moments are computed in terms of the magnetic moments of their constituent quarks. For example, for $\Omega_{cb}^0$ with $J=1/2$ one has $\mu_B = (2\mu_b + 2\mu_c - \mu_s)/3$ and for $\Xi_{cb}^+$ it reads $\mu_B = (2\mu_b + 2\mu_c - \mu_u)/3$. Moreover, for $\Sigma_{cb}^+$ and $\Omega_{cb}^0$ with $J=3/2$ one has $\mu_B = \mu_b + \mu_c + \mu_u$ and $\mu_B = \mu_b + \mu_c + \mu_s$, respectively. As is expected, the magnetic moments of baryons containing two heavy quarks are dominated by the contribution of the light quark, since the magnetic moment is inversely proportional to the quark mass. Consequently, baryons with identical light-quark content but different heavy flavors exhibit similar magnetic moments. This behavior is clearly observed, for example, in the comparison between the $\Xi_{cc}$ and $\Xi_{bb}$ states. For the spin-3/2 baryons, the magnetic moments are generally larger than those of their spin-1/2 counterparts due to the aligned spin configuration of the constituent quarks. From these Tables, it is also observed that our results are in reasonable agreement with previous quark model and light-cone QCD sum rule predictions [27,34,57,60], although some discrepancies are observed in specific channels. These differences mainly originate from the choice of quark masses, the treatment of relativistic corrections, and the spin–flavor structure adopted in each model. Experimental measurements of the magnetic moments of heavy flavor baryons are sparse and few experimental groups such as BTeV and SELEX Collaborations [62] are expected to do these measurements in the near future.

Table 4. Magnetic moments of $\Xi$ in units of nuclear magneton.

| Baryon | $J^P$ | $\mu$ | $\mu$ [57] | Baryon | $J^P$ | $\mu$ | $\mu$ [other] |
|---|---|---|---|---|---|---|---|
| $\Xi_{cc}^{++}$ | $1/2^+$ | $0.17 \pm 0.05$ | $0.13^{+0.52}_{-0.38}$ | $\Xi_{cb}^{\prime 0}$ | $1/2^+$ | $0.53 \pm 0.13$ | $0.42^{+0.24}_{-0.18}$ [57] |
| $\Xi_{cc}^+$ | $1/2^+$ | $0.84 \pm 0.13$ | $0.72^{+0.52}_{-0.20}$ | $\Xi_{cb}^{*+}$ | $3/2^+$ | $2.34 \pm 0.24$ | $2.27^{+0.27}_{-0.14}$ [60] |
| $\Xi_{bb}^0$ | $1/2^+$ | $-0.59 \pm 0.09$ | $-0.53^{+0.06}_{-0.47}$ | $\Xi_{cb}^{*0}$ | $3/2^+$ | $-0.46 \pm 0.08$ | -0.39 [27] |
| $\Xi_{bb}^-$ | $1/2^+$ | $0.23 \pm 0.08$ | $0.18^{+0.24}_{-0.06}$ | $\Xi_{cc}^{*++}$ | $3/2^+$ | $2.74 \pm 0.34$ | $2.67^{+0.27}_{-0.15}$ [60] |
| $\Xi_{cb}^+$ | $1/2^+$ | $1.61 \pm 0.23$ | $1.52^{+0.002}_{-1.52}$ | $\Xi_{bb}^{*0}$ | $3/2^+$ | $1.47 \pm 0.19$ | 1.37 [27] |
| $\Xi_{cb}^0$ | $1/2^+$ | $-0.79 \pm 0.09$ | $-0.76^{+0.002}_{-0.76}$ | $\Xi_{bb}^{*-}$ | $3/2^+$ | $-1.20 \pm 0.16$ | $-1.11^{+0.06}_{-0.14}$ [60] |
| $\Xi_{cb}^{\prime +}$ | $1/2^+$ | $-0.14 \pm 0.04$ | $-0.12^{+0.24}_{-0.36}$ | $\Xi_{cc}^{*+}$ | $3/2^+$ | $-0.39 \pm 0.09$ | $-0.311^{+0.052}_{-0.130}$ [60] |

Table 5. Magnetic moments of Ω in units of nuclear magneton.

| Baryon | J | μ | μ [other] | Baryon | J | μ | μ [other] |
|---|---|---|---|---|---|---|---|
| $\Omega_{cc}^{+}$ | $1/2^{+}$ | $0.711 \pm 0.191$ | $0.67_{-0.14}^{+0.53}$ [57] | $\Omega_{cbb}^{+}$ | $1/2^{+}$ | $-0.231 \pm 0.061$ | $-0.199$ [34] |
| $\Omega_{bb}^{-}$ | $1/2^{+}$ | $0.070 \pm 0.020$ | $0.04_{-0.08}^{+0.12}$ [57] | $\Omega_{bbb}^{-}$ | $1/2^{+}$ | $-0.202 \pm 0.062$ | $-0.192$ [34] |
| $\Omega_{cb}^{0}$ | $1/2^{+}$ | $-0.691 \pm 0.182$ | $-0.61_{-0.61}^{+0.002}$ [57] | $\Omega_{cc}^{*+}$ | $3/2^{+}$ | $0.430 \pm 0.093$ | $0.390$ [27] |
| $\Omega_{cb}'^{0}$ | $1/2^{+}$ | $0.521 \pm 0.112$ | $0.450_{-0.20}^{+0.25}$ [57] | $\Omega_{bb}^{*-}$ | $3/2^{+}$ | $-0.662 \pm 0.171$ | $-0.662_{-0.024}^{+0.022}$ [60] |
| $\Omega_{ccb}^{+}$ | $1/2^{+}$ | $0.572 \pm 0.130$ | $0.492$ [34] | $\Omega_{cb}^{*0}$ | $3/2^{+}$ | $-0.270 \pm 0.073$ | $-0.261_{-0.021}^{+0.015}$ [60] |

As was mentioned, radiative decays of doubly heavy baryons provide a sensitive probe into the magnetic moments of heavy quarks in a multiquark environment. Here, considering the calculated transition magnetic moments, by the help of Eq. (2) we evaluate the radiative decay widths of doubly heavy flavored baryons for the radiative transitions of the type $B(J=3/2) \to B(J=1/2) + \gamma$. Results are very sensitive to the mass difference of initial and final heavy baryons. In fact, if the mass of one of the baryons changes even at mili-digit level, the decay width can change several times. Results are listed in Table 6 and compared with theoretical predictions reported in Refs. [35, 63-68]. Our results indicate that radiative decay widths are generally small, ranging from fractions of keV to a few keV, which is characteristic of electromagnetic transitions in heavy baryons. Among the studied channels, the $\Xi_{cc}^{*} \to \Xi_{cc}\gamma$ and $\Omega_{cc}^{*} \to \Omega_{cc}\gamma$ transitions exhibit relatively larger widths, reflecting the stronger contribution of the lighter charm quark compared to bottom quarks. For bottom and bottom–charm baryons, the decay widths are significantly suppressed due to the large masses of the bottom quarks, leading to smaller magnetic moments and reduced phase space. Our predictions show good agreement with other quark model calculations such as the framework of the modified bag model [35] or the relativistic constituent three–quark model [63], with minor deviations that can be traced back to differences in baryon mass splittings and transition magnetic moments $\mu(B^{*} \leftrightarrow B)$.

Table 6. Radiative decay width for double heavy baryons ($\Gamma(B(J=3/2) \to B(J=1/2)\gamma)$) (in keV).

| Decay mode | Our | Others | Decay mode | Our | Others |
|---|---|---|---|---|---|
| $\Gamma(\Xi_{cc}^{*++} \to \Xi_{cc}^{++}\gamma)$ | $1.540 \pm 0.181$ | 1.43 [35]<br>3.90 [65]<br>14.6 [67]<br>1.64 [68] | $\Gamma(\Omega_{cc}^{*+} \to \Omega_{cc}^{+}\gamma)$ | $0.962 \pm 0.152$ | 0.949 [35]<br>0.326 [68] |
| $\Gamma(\Xi_{bb}^{*0} \to \Xi_{bb}^{0}\gamma)$ | $0.432 \pm 0.102$ | 0.31 [63]<br>0.022 [66]<br>0.001 [68] | $\Gamma(\Omega_{bb}^{*} \to \Omega_{bb}\gamma)$ | $0.061 \pm 0.011$ | 0.02 [63]<br>0.04 [65]<br>0.001 [68] |
| $\Gamma(\Xi_{bc}^{*+} \to \Xi_{bc}^{+}\gamma)$ | $0.251 \pm 0.082$ | 0.209 [64]<br>0.612 [66] | $\Gamma(\Omega_{bc}^{*0} \to \Omega_{bc}^{0}\gamma)$ | $0.0054 \pm 0.0009$ | 0.003 [64]<br>0.002 [68] |

# 6. Conclusion

Heavy baryons form unique laboratories to study the dynamics of strong interaction as well as color confinement in the theory of QCD. They also provide an excellent laboratory to understand the dynamics of light flavor quarks in the vicinity of heavy flavor quarks. Doubly heavy baryons are particularly interesting because they offer a new platform for exploring heavy quark symmetry and chiral dynamics, simultaneously. Regarding recent results from the CERN LHCb on the singly and doubly charmed baryons there is renewed interest both experimentally and theoretically in the static properties of heavy flavor baryons such as their masses, magnetic moments and radiative decay widths.

From the theoretical point of view, the mass spectra and the electromagnetic radiative decay of bottom flavored hadrons have been computed by various approaches such as the leading-order heavy quark effective theory, chiral SU(3) quark model, heavy quark and chiral symmetries, light cone QCD sum rules, etc. In spite of all efforts, in one hand, there exist wide disparities among the predicted values of heavy baryons properties and, on the other hand, many expected triply heavy flavored baryons have not yet observed experimentally and the quantum status of some observed doubly heavy baryons has not also determined accurately. This motivated us to look for the several aspects of the constituent quark model with heavy flavor contents and to invoke different model descriptions of the basic quark-quark interaction potential. In this work, to study the properties of heavy baryons we used a theoretical approach based on the approximative quark-diquark model. It is proven that the simple diquark model is very successful in describing many interesting phenomena. This allows us to convert a three-body problem into a two-body one so that the Bethe–Salpeter equation, convenient for two-body systems, can be applied. Through the quark-diquark model, we computed the S-wave mass spectra of doubly and triply heavy baryons as well as their magnetic moments and radiative decay widths. The magnetic moments obtained in our study highlight the dominant contribution of the light quark to the electromagnetic properties of doubly heavy baryons, while the heavy diquark contribution is suppressed by its large mass. Our results have been compared with the theoretical models as well as existing experimental data so that good agreements were found with other theoretical approaches and the only available experimental data for the doubly charmed sector. The overall patterns and magnitudes of the magnetic moments are compatible with expectations from heavy-quark symmetry and are in reasonable agreement with previous model predictions. We computed the radiative decay widths for the $J=3/2 \rightarrow J=1/2$ transitions which are sensitive to the underlying spin structure. Our results exhibit trends consistent with other theoretical estimates. Observed consistencies support the validity of the dynamical heavy diquark model in describing the internal structure of heavy baryons as well as our description of the quark-quark interaction potential.

It is expected that heavy baryons will be copiously produced in future colliders such as electron–positron international linear collider (ILC) and circular electron–positron collider (CEPC). These permit us to test our predictions for radiative decays and, hence, to determine the size of corrections to these leading order evaluations. Moreover, the predictions presented in this work may serve as useful benchmarks for lattice QCD simulations, particularly in the bottom and bottom–charm sectors where experimental information is still scarce.

**Data Availability Statement** This manuscript has no associated data or the data will not be deposited. [Authors' comment: The data used in our work could be found in the references introduced in our work.]

# Appendix

In this section, the spin-flavor combinations and their wave functions related to the spin-1/2 and spin-3/2 heavy baryons, suitable to be used in Eq. (1), are given.

**Table I.** Spin-flavor wave functions of heavy flavor baryons with $J^P = (1/2)^+$

| Baryon | Spin-flavor wave function |
|---|---|
| $\Xi_{cc}^+$ | $\frac{\sqrt{2}}{6}(2d_-c_+c_+ - c_-d_+c_+ - d_+c_-c_+ + 2c_+d_-c_+ - c_+c_-d_+ - c_-c_+d_+ - c_+d_+c_- - d_+c_+c_- + 2c_+c_+d_-)$ |
| $\Xi_{cc}^{++}$ | $\frac{\sqrt{2}}{6}(2u_-c_+c_+ - c_-u_+c_+ - u_+c_-c_+ + 2c_+u_-c_+ - c_+c_-u_+ - c_-c_+u_+ - c_+u_+c_- - u_+c_+c_- + 2c_+c_+u_-)$ |
| $\Xi_{bb}^-$ | $\frac{\sqrt{2}}{6}(2d_-b_+b_+ - b_-d_+b_+ - d_+b_-b_+ + 2b_+d_-b_+ - b_+b_-d_+ - b_-b_+d_+ - b_+d_+b_- - d_+b_+b_- + 2b_+b_+d_-)$ |
| $\Xi_{bb}^0$ | $\frac{\sqrt{2}}{6}(2u_-b_+b_+ - b_-u_+b_+ - u_+b_-b_+ + 2b_+u_-b_+ - b_+b_-u_+ - b_-b_+u_+ - b_+u_+b_- - u+b_+b_- + 2b_+b_+u_-)$ |
| $\Omega_{bb}^-$ | $\frac{\sqrt{2}}{6}(2s_-b_+b_+ - b_-s_+b_+ - s_+b_-b_+ + 2b_+s_-b_+ - b_+b_-s_+ - b_-b_+s_+ - b_+s_+b_- - s_+b_+b_- + 2b_+b_+s_-)$ |
| $\Omega_{cc}^+$ | $\frac{\sqrt{2}}{6}(2s_-c_+c_+ - c_-s_+c_+ - s_+c_-c_+ + 2c_+s_-c_+ - c_+c_-s_+ - c_-c_+s_+ - c_+s_+c_- - s_+c_+c_- + 2c_+c_+s_-)$ |
| $\Omega_{cb}^0$ | $\frac{-1}{6}(b_+c_-s_+ + c_+b_-s_+ + s_+c_-b_+ + s_+b_-c_+ - 2b_+s_-c_+ - 2c_+s_-b_+ + b_-c_+s_+ + c_-b_+s_+ - 2s_-c_+b_+$ $-2s_-b_+c_+ + b_-s_+c_+ + c_-s_+b_+ - 2b_+c_+s_- - 2c_+b_+s_- + s_+c_+b_- + s_+b_+c_- + b_+s_+c_- + c_+s_+b_-)$ |
| $\Xi_{cb}^0$ | $\frac{-1}{6}(b_+c_-d_+ + c_+b_-d_+ + d_+c_-b_+ + d_+b_-c_+ - 2b_+d_-c_+ - 2c_+d_-b_+ + b_-c_+d_+ + c_-b_+d_+ - 2d_-c_+b_+$ $-2d_-b_+c_+ + b_-d_+c_+ + c_-d_+b_+ - 2b_+c_+d_- - 2c_+b_+d_- + d_+c_+b_- + d_+b_+c_- + b_+d_+c_- + c_+d_+b_-)$ |
| $\Xi_{cb}^+$ | $\frac{-1}{6}(b_+c_-u_+ + c_+b_-u_+ + u_+c_-b_+ + u_+b_-c_+ - 2b_+u_-c_+ - 2c_+u_-b_+ + b_-c_+u_+ + c_-b_+u_+ - 2u_-c_+b_+$ $-2u_-b_+c_+ + b_-u_+c_+ + c_-u_+b_+ - 2b_+c_+u_- - 2c_+b_+u_- + u_+c_+b_- + u_+b_+c_- + b_+u_+c_- + c_+u_+b_-)$ |
| $\Omega_{ccb}^+$ | $\frac{\sqrt{2}}{6}(2b_-c_+c_+ - c_-b_+c_+ - b_+c_-c_+ + 2c_+b_-c_+ - c_+c_-b_+ - c_-c_+b_+ - c_+b_+c_- - b_+c_+c_- + 2c_+c_+b_-)$ |
| $\Omega_{cbb}^0$ | $\frac{\sqrt{2}}{6}(2c_-b_+b_+ - b_-c_+b_+ - c_+b_-b_+ + 2b_+c_-b_+ - b_+b_-c_+ - b_-b_+c_+ - b_+c_+b_- - c_+b_+b_- + 2b_+b_+c_-)$ |

**Table II.** Spin-flavor wave functions of heavy flavor baryons with $J^P = (3/2)^+$

| Baryons | Spin-flavor wave functions |
|---|---|
| $\Xi_{cc}^+$ | $\frac{1}{\sqrt{3}}(c_+c_+d_+ + c_+d_+c_+ + d_+c_+c_+)$ |
| $\Xi_{cc}^{++}$ | $\frac{1}{\sqrt{3}}(c_+c_+u_+ + c_+u_+c_+ + u_+c_+c_+)$ |
| $\Xi_{bb}^+$ | $\frac{1}{\sqrt{3}}(b_+b_+d_+ + b_+d_+b_+ + d_+b_+b_+)$ |
| $\Xi_{bb}^{++}$ | $\frac{1}{\sqrt{3}}(b_+b_+u_+ + b_+u_+b_+ + u_+b_+b_+)$ |
| $\Omega_{bb}^{-}$ | $\frac{1}{\sqrt{3}}(b_+b_+s_+ + b_+s_+b_+ + s_+b_+b_+)$ |
| $\Omega_{cc}^{+}$ | $\frac{1}{\sqrt{3}}(c_+c_+s_+ + c_+s_+c_+ + s_+c_+c_+)$ |
| $\Omega_{bc}^{0}$ | $\frac{1}{\sqrt{6}}(b_+c_+s_+ + c_+b_+s_+ + b_+s_+c_+ + s_+b_+c_+ + c_+s_+b_+ + s_+c_+b_+)$ |
| $\Xi_{bc}^{0}$ | $\frac{1}{\sqrt{6}}(b_+c_+d_+ + c_+b_+d_+ + b_+d_+c_+ + d_+b_+c_+ + c_+d_+b_+ + d_+c_+b_+)$ |
| $\Xi_{bc}^{+}$ | $\frac{1}{\sqrt{6}}(b_+c_+u_+ + c_+b_+u_+ + b_+u_+c_+ + u_+b_+c_+ + c_+u_+b_+ + u_+c_+b_+)$ |
| $\Omega_{bbb}^{*-}$ | $b_+b_+b_+$ |